\documentclass[12pt]{article}
\usepackage{amsmath}
\usepackage{graphicx}
\usepackage{amssymb}
\usepackage{amsfonts}
\usepackage{latexsym}
\usepackage[latin1]{inputenc}

\renewcommand{\vec}[1]{{\bf #1}}

\def\beq{\begin{eqnarray}}
\def\eeq{\end{eqnarray}}

\def\ln{\,\mbox{ln}\,}

\def\al{\alpha}
\def\be{\beta}

\def\de{\delta}
\def\vp{\varepsilon}

\def\ka{\kappa}

\def\ph{\varphi}

\def\Om{\Omega}

\begin{document}

\begin{center}
{\large \sc Dynamics of the Laplace-Runge-Lenz vector in the
quantum-corrected Newton gravity}
\vskip 6mm

{\bf C. Farina$^a$, W. J. M. Kort-Kamp$^a$, Sebastiao Mauro$^b$,
Ilya L. Shapiro$^{b}$\footnote{Also at Tomsk State Pedagogical University,
 Russia. \ E-mail address: shapiro@fisica.ufjf.br} }
\vskip 6mm

{\small\sl (a) \ Instituto de F\'{\i}sica,
Universidade Federal de Rio de Janeiro}
\vskip 2mm

{\small\sl (b) \ Departamento de F\'{\i}sica -- ICE,
Universidade Federal de Juiz de Fora }

 {\small\sl Juiz de Fora, CEP: 36036-330, MG,  Brazil}
\end{center}

\vskip 10mm

\begin{quotation}
\centerline{\large\bf Abstract}
\vskip 2mm
\noindent
Recently it was shown that quantum corrections to the Newton
potential can explain the rotation curves in spiral galaxies
without introducing the Dark Matter halo. The unique
phenomenological parameter $\al\nu$ of the theory grows
with the mass of the galaxy. In order to better investigate
the mass-dependence of $\al\nu$ one needs to check the
upper bound for $\al\nu$ at a smaller scale. Here we perform
the corresponding calculation by analyzing the dynamics of 
the Laplace-Runge-Lenz vector. The resulting limitation on  
quantum corrections is quite severe, suggesting a strong 
mass-dependence of $\al\nu$.
\end{quotation}
\vskip 6mm

\section{\large\bf Introduction}

It is a common belief nowadays that the General Relativity (GR)
is not the ultimate theory of gravity. One reason for this is
that the relevant solutions of GR, such as the spherically
symmetric solution and the homogeneous and isotropic one, both
manifest singular behavior in their extremes. In the first
case the space-time singularity is in the center of the
black hole and in the second case it is in the initial instant
of the universe ``Big Bang''. In both cases the singularity
is surrounded by a very small space-time region with very
high magnitudes of curvature tensor components. This makes
perfectly possible
that the higher derivative terms in the gravitational action
may change the geometry in such a way that the singularities
should disappear. The importance of higher derivative terms
in the gravitational action is due to the fact that they are
requested for constructing a renormalizable theory of matter
(including Standard Model) on curved background (see, e.g.,
\cite{book} and further references therein). The effects of
higher derivative terms on singularities were discussed in
the cosmological (see, e.g., \cite{star,ander,anju}) and
black hole (see, e.g., \cite{FroVil-80,BH-HD}) settings and
there are serious reasons to consider the possibility of
erased singularity due to the higher derivative terms in the
classical action and quantum corrections \cite{PoImpo}.

Despite the discussion of higher derivative terms and their
possible effect on singularities is interesting, we are
much more curious about possible modifications of gravity
and, especially, about possible quantum effects at low
energy scale, where the observations are much more real.
At low energies the effect of higher derivative terms is
usually assumed to be Planck-suppressed and one has to deal
with, e.g., quantum corrections to the Hilbert-Einstein
action. For the higher derivative section of the theory, direct
calculations of the low-energy quantum contributions were recently
performed in \cite{apco,fervi} using the approximation of
linearized metric on flat background. At the same time it was
demonstrated in \cite{apco} that this kind of calculational
technique is useless for deriving quantum corrections for the
cosmological and Hilbert-Einstein terms. Therefore, despite this
part is much more interesting from the physical point of view, 
here we have much less achievements. As far as the subject 
looks quite relevant for applications, it is worthwhile to 
try some phenomenological approach in this case.

In what follows we will consider the low-energy quantum
corrections to the Newton potential using the renormalization
group technique and the identification of renormalization
scale which was recently proposed in \cite{RotCurve} on the
phenomenological basis and then justified theoretically in
\cite{Hrvoje}. This identification of scale includes some
uncertainty which is measured by a dimensionless 
phenomenological parameter $\al\nu$. It was shown in 
\cite{RotCurve} that values of $\al\nu$ of the order 
$10^{-7}$ can provide the detailed and precise
explanation for the rotation curves for a small but quite
representative sample of spiral galaxies. It is remarkable
that the mentioned parameter is steadily growing with the
increase of the mass of the galaxy. Therefore, it is
natural to expect a much smaller value for $\al\nu$
for much smaller astrophysical systems, such 
as starts. Our purpose is to set an upper bound on 
$\al\nu$ for the Solar system, by using a very efficient 
approach based on the dynamics of the Laplace-Runge-Lentz 
vector (see, for instance, \cite{Sivardiere84} and references 
therein for the introduction, for an interesting historical 
review, see \cite{Goldstein75}).

The paper is organized as follows. In the next section we
present very general arguments about unique possible form
of renormalization group running of Newton constant $G$
in the IR region. In Sect. 3 a brief introduction and
necessary information about the  Laplace-Runge-Lentz vector
in the almost Keplerian problem are presented. The
numerical estimates and the upper bound for $\al\nu$
from the Mercury precession are derived in Sect. 4. In
Sect. 5 we draw our conclusions.

\section{\large\bf Quantum effects and Newton potential}

One of the most powerful techniques for evaluating quantum
corrections is renormalization group. So, let us check out
what is the renormalization group equation in the low-energy
gravitational sector. The unique relevant parameter at the
astrophysical scale is the Newton constant $G$, and at the
quantum level it becomes a running parameter $G(\mu)$, where
$\mu$ defines a scale. The problem of identifying $\mu$
with some physical quantity will be discussed later on,
and now we concentrate on the dependence  $G(\mu)$, which
is always governed by the corresponding renormalization group
equation \ $\mu (dG/d\mu) = \be_G$.

Consider an {\it arbitrary} quantum theory with gravity.
It can be, for instance, some quantum theory of the
gravitational field or quantum theory of matter fields.
Every kind of quantum theory can be characterized by the
massive parameters $m$ which define scale. 
There may be, of course, more than one such parameter, 
so let us consider, for the sake of generality, the 
whole set $\{ m_i\}$. For example, the elements of the set 
$\{ m_i\}$ can be the masses of all particles 
or fields which are present in the given quantum theory.

Let us present general arguments about the possible
form of the running Newton constant, $G(\mu)$.
Using dimensional arguments we can establish the
unique possible form of the renormalization group equation,
\beq
\mu \frac{dG^{-1}}{d\mu}\,=\,\sum\limits_{particles}
\,A_{ij}\,m_i\,m_j\,=\,2\nu\,M_P^2 \,,\qquad
G^{-1}(\mu_0)=G^{-1}_0=M^2_P\,.
\label{RG1}
\eeq
In particular, in the SM-like or GUT-like 
theory, at the one-loop level, one has
\beq
\sum\limits_{particles}A_{ij}\,m_i\,m_j
\,=\,\,\sum\limits_{fermions} \frac{m_f^2}{3(4\pi)^2}
\,-\,\sum\limits_{scalars} \frac{m_s^2}{(4\pi)^2}
\Big(\xi_s-\frac16\Big)\,,
\label{RG2}
\eeq
where the fermion masses were denoted by $m_f$ and $\xi_s$ 
is the nonminimal parameter for the scalar with the mass $m_s$.

At the same time, it is important to stress that Eq.
(\ref{RG1}) is {\it not} just a one-loop equation, but it
is valid at {\it any} loop order and in any theory which is
capable to produce the renormalization group equation for $G$.
In general, beyond the one-loop level, the coefficients 
$A_{ij}$ in Eq.  (\ref{RG1}) depend on coupling constants 
which are present in the theory. 

One can rewrite Eq. (\ref{RG1}) as
\beq
\mu \frac{d(G/G_0)}{d\mu}\,=\,-\,2\nu\,(G/G_0)^2 \,.
\label{RG3}
\eeq
Solving this equation we obtain the universal form of the
scale dependence for the Newton constant.
\beq
G(\mu)=\frac{G_0}{1 + \nu\,\ln\left(\mu^2/\mu_0^2\right)}\,.
\label{RG4}
\eeq
Here the word \ {\it universal} \ means that either $G$ is
not running (and this means there are no quantum effects in
the low-energy gravity sector) or such running is given by
Eq. (\ref{RG4}). From this perspective it is not a surprise
that Eqs. (\ref{RG3}) and (\ref{RG4}) can be met in
such qualitatively different approaches as higher derivative
quantum gravity \cite{SalStr,frts82,AvrBar,Gauss},
quantum theory of matter fields on curved background
\cite{nelspan82,book} and quantum theory of conformal factor
\cite{AntMot92}. The same equation (\ref{RG4})
shows up also in the phenomenological approach based
on the hypothesis of the Appelquist and Carazzone - like
decoupling for the cosmological constant and conservation
law for the quantum corrected gravitational action
\cite{Gruni}. The reason behind these occurrences is
that any other form of $G(\mu)$ would be in conflict with
very simple (and hence very safe) dimensional considerations
and also with the covariance arguments which play a very
significant role here \cite{PoImpo}.

The next problem is how to identify $\mu$. As usual, this
identification depends on the physical problem under
discussion and there is no universal solution. In the case
of the gravitational field of a point-like mass, the most
natural choice is \ $\mu \ \sim 1/r$, \ where $r$ is the
distance from the mass position. This choice of the scale
identification has been used in various publications
\cite{{Bertol},{Eli-94},{Reuter:2007de},{Gruni}}. In particular,
this identification enables one to roughly explain the flat
rotation curves for the point-like model of the spiral galaxies
\cite{{Bertol},{Eli-94},{Gruni}} by directly using Eq. (\ref{RG4}).
Moreover, one can achieve very good and detailed description 
for the good sample of rotation curves by using Eq. (\ref{RG4})
together with more sophisticated identification of the scale,
\beq
\frac {\mu}{\mu_0}
=  \Big( \frac{\Phi_{\mbox{\tiny Newt}}}{\Phi_0} \Big)^\alpha\,,
\label{muset}
\eeq
where the value of $\Phi_0$ is irrelevant and $\Phi_{\rm Newt}$ is
the Newtonian potential computed with the boundary condition
of it being zero at infinity. In Eq. (\ref{muset}), $\alpha$ 
is a phenomenological parameter which should be defined from 
fitting to the
observational data. Let us note that the same dependence
(\ref{muset}) can be also obtained from the regular
scale-setting procedure \cite{Guberina-scale}, which was
applied to the present case in \cite{Hrvoje}.

It is easy to see from Eqs. (\ref{RG4})
and (\ref{muset}) that $\al$ always shows up as a factor in
the product \ $\al\nu$. It turns out that the fit with the
observational data is perfect (for a sample of nine galaxies)
if we assume that the product \ $\al\nu$ \ is about $10^{-7}$
and,  moreover, this product grows up with the increasing 
of the mass of the galaxy \cite{RotCurve}.
Indeed, this is a very nice feature, because then one may
hope that the effect of ``corrected'' Newton law would be
very weak at the scale of the Solar system, which has a
mass of many orders of magnitude smaller than the one of
a galaxy. The purpose of the present paper is to make the
last statement quantitative, that is to set an upper bound
on the value of  \ $\al\nu$ \  inside the Solar system.
The best available data here are about the precession of the
perihelion of Mercury, so it is sufficient to deal with these
data only.

\section{\large\bf  Laplace-Runge-Lenz vector}

The method that we will use to calculate the precession in the
orbit of the Mercury, due to quantum effect in the Newtonian
potential, is based on the known Laplace-Runge-Lenz vector.
Therefore we shall do a discuss briefly about it and we will
see as we can use it to calculate the precession velocity.

Consider first the motion of a particle of mass $m$ for the
non-perturbed Kepler's problem, when there is a single Newtonian
force acting on the particle,
\beq
{\bf F}_{Newt} = -\dfrac{k}{r^2}\hat{{\bf r}} \, ,
\eeq
where $k = G_0Mm$. The Laplace-Runge-Lenz vector (LRL) 
is defined as
\beq
{\bf A} \,=\, {\bf p}\times
\mbox{{\mathversion{bold}${\ell}$}} -
mk{\bf {\hat r}}\, ,
\eeq
where 
$\mbox{{\mathversion{bold}${\ell}$}}={\vec r} \times {\vec p}$
is the angular momentum vector.
One can show that in the case of the non-perturbed Kepler's
problem LRL is a constant of motion, namely,
\beq
\dfrac{d{\bf A}}{dt} = \dfrac{d{\bf p}}{dt}\times
\mbox{{\mathversion{bold}${\ell}$}} +
 {\bf p}\times\dfrac{d\mbox{{\mathversion{bold}${\ell}$}}}{dt}
 - mk\dfrac{d{\bf {\hat r}}}{dt}\ = {\bf 0}\,.
\eeq
Furthermore, LRL has some important relationships with other
constants of motion. For example,
$\mbox{{\mathversion{bold}${\ell}$}} \cdot{\bf A} = 0$,
hence LRL always remains in the plane of the orbit. Another
important relation concerns the total energy of the particle
$E$ and to its angular momentum
\beq
\label{AQuadrado}
{\bf A}^2 &=& ({\bf p}\times \mbox{{\mathversion{bold}${\ell}$}} -
mk{\bf {\hat r}})\cdot ({\bf p}\times
\mbox{{\mathversion{bold}${\ell}$}} - mk{\bf {\hat r}})
\,=\,  m^2k^2\left(
1+\dfrac{2E\mbox{{\mathversion{bold}${\ell}$}}^2}{mk^2}\right)\,.
\eeq
remembering that for the Kepler problem the eccentricity of 
the orbit is related to the energy and angular momentum as
\beq
\label{excentricidade}
\varepsilon = \sqrt{1
+\dfrac{2E\mbox{{\mathversion{bold}${\ell}$}}^2}{mk^2}} \, ,
\eeq
so that we can write the modulo of the LRL vector as
\beq
|{\bf A}| = mk\varepsilon \,.
\eeq
One can see that the magnitude of the LRL vector measures 
the eccentricity of the orbit. Moreover, one can arrive at
the equation of the orbit by taking scalar product of the LRL
with the position vector, namely
\beq
r\vert{\bf A}\vert\,\mbox{cos}(\varphi-\varphi_0) =
 {\bf r}\cdot\Bigl({\bf p}\times
\mbox{{\mathversion{bold}${\ell}$}} -
mk{\bf{\hat r}}\Bigr) =
 \ell^2 - mkr\,,
\eeq
where  $\varphi_0$ is the angle between ${\bf A}$ and polar
axis. After some simple manipulations we get
\beq
 \label{EqConica}
 r = \frac{\ell^2/mk}{1 +
  \mbox{\Large $\frac{\vert{\bf A}\vert}{mk}$}
\,\cos (\varphi - \varphi_0)}\, .
\eeq
One can see that ${\bf A}$ is pointing to the direction
of symmetry of the orbit, which can be limited or unlimited.
Then it is convenient to choose the polar axis in the direction
of the LRL vector ${\bf A}$, e.g., by choosing $\varphi_0=0$.
Then ${\bf A} = mk\varepsilon \hat{{\bf x}}$ and
\beq
r = \frac{\ell^2/mk}{1 +
\varepsilon\,\mbox{cos}\varphi}
= \frac{a(1-\varepsilon^2)}{1 +
  \varepsilon\,\mbox{cos}\varphi}\, .
\label{EqOrbita}
\eeq
where $a$ is the major semi-axis of the ellipse.

Once we know the properties and the interpretation of the
LRL vector, we are able to deal with the Kepler's problem when  
a small perturbation is introduced. In this case the new orbit
will be very similar to the old one, however there will be a 
precession. In other words, the particle has an approximately 
elliptical orbit, but in such a way that the major semi-axis 
slowly rotates. The velocity of this rotation is called 
precession velocity. 
A comparison between theoretical predictions and available 
experimental (observational) data may give valuable information 
such as, for instance, upper bounds on relevant parameters 
which are present in the perturbing force. In our case, we 
shall follow this line by using the well-known data for the 
Mercury precession. 

Consider a particle of mass $m$ which moves under the action
of a total force
\beq
{\bf F} = - \frac{k}{r^2}\,{\bf{\hat r}} + {\bf f}\,.
\eeq
Here $\,{\bf f}$ is a small perturbation force
($|{\bf f}|\ll k/r^2$), which can be non-central, in principle. 
Some formulas below are general, but in fact we are mainly 
interested in the central perturbed force
$\,{\bf f} = - \,{\rm grad}\,u(r)$, where the perturbing term 
for the potential energy can be derived from the Eqs. (\ref{RG4}) 
and (\ref{muset}) to be \cite{Gruni,RotCurve}
\beq
\Phi = \Phi_{\rm Newt} + u(r)\,, \qquad
u(r) \,=\,\frac{mc^2}{2} \,\, \frac{\de G}{G_0}\,,
\label{Phi}
\eeq
where \ $\delta G = G(\mu) - G_0$.

Let us present, for the sake of completeness, the short review
of the derivation of Eq. (\ref{Phi}). One can start from the 
expression for the action with variable $G$, 
\beq
S_{grav} \,=\, \frac{1}{16\pi} 
\int d^4x\sqrt{-g}\,\frac{R}{G(\mu)}\,,
\label{varG}
\eeq
where $\mu$ is supposed to depend on some energy features of 
the gravitational field. We assume that $G$ has a very weak 
deviation from the constant value $G_0=1/M^2_{Pl}$, namely 
$G=G_0(1+\ka)$. In what follows we will consider $\ka$ to 
be a very small quantity and hence keep only first order
terms in this parameter. Our purpose is to link the action 
(\ref{varG}) with the usual one of GR, with $G_0$ instead 
of $G(\mu)$. For this end we perform the conformal 
transformation \cite{Gruni} according to 
\beq
{\bar g}_{\mu \nu} = \frac{G_0}{G} ~ g_{\mu \nu}
= (1-\ka)g_{\mu \nu}\,. 
\label{conf}
\eeq
The derivatives of $\ka$ emerge in the transformed action, 
but only in second power, hence they may be neglected. Then, 
in the linear order in $\ka$, the metric ${\bar g}_{\mu \nu}$ 
satisfies Einstein equations with constant $G_0$, and the 
nonrelativistic limit of the two metrics 
${\bar g}_{\mu \nu}$ and $g_{\mu \nu}$ is related as 
\beq
g_{00} = - 1 -  \frac {2 \Phi}{c^2} \,, \quad 
\mbox{hence} \quad 
{\bar g}_{00} = - 1 - \frac{2 \Phi_{\rm Newt}}{c^2}\,.
\label{00}
\eeq
Here $\Phi_{\rm Newt}$ is the usual Newton potential and 
$\Phi$ is an apparent potential corresponding to the 
nonrelativistic solution of the modifies gravitational 
theory (\ref{varG}). Due to the (\ref{00}), we arrive at 
\cite{Gruni,RotCurve}
\beq
\Phi = \Phi_{\rm Newt} + \frac{c^2} 2\,\ka 
= \Phi_{\rm Newt} + \frac{c^2 \, \de G}{2\,G_0}\,,
\label{Phi new}
\eeq
which is nothing else but (\ref{Phi}). Now we can come back to 
the analysis of this formula. For the Solar system case we 
can use the relation $\,\Phi_{\rm Newt}=-k/r$. Hence, the 
identification (\ref{muset}) effectively coincides with the 
\ $\mu \sim 1/r$ \ one of the Refs. 
\cite{Bertol,Eli-94,Reuter:2007de,Gruni}.

One can easily prove the following statement: For the case of
a central perturbation force $\,{\bf f} = - \,{\rm grad}\,u(r)$,
the magnitude of the LRL vector varies according to the relation
\beq
d A^2\, = \, -2 m\,\ell^2\,\,d u(r) =
\, -2 m\,\ell^2\,u^\prime(r)\, dr\,.
\label{5}
\eeq

The last relation shows that the magnitude of the vector
${\vec A}$ varies as the distance $r$ varies, but in such 
a way that it takes equal values for equal values of $r$. 
If we restrict the discussion by quase-elliptic orbits, 
which are restricted in space, we have 
$r_1 \leq r \leq r_2$, where $r_1$ and $r_2$ are the 
corresponding turning points,
then the modulus of the Laplace-Runge-Lenz vector  
${\vec A}$ will assume the same value whenever the 
distance $r$ is the same. Hence, the dynamics of ${\vec A}$ 
is perfectly useful for evaluating the precession of the 
orbit in the quasi-Newtonian case. The variation 
of the magnitude $|{\vec A}|$ is of the first order of 
magnitude in the small perturbation $u(r)$. Therefore, we 
can completely neglect this variation when evaluating the 
precession of the LRL vector, because this precession is 
also of the first order in $u(r)$. In what follows we will 
assume, for the sake of simplicity, that $|{\vec A}|$ is 
constant even when the small perturbations are present.

The time variation rate of the LRL vector is given by
\beq
\label{LRLDifZero}
\frac{d{\bf A}}{dt} \,\,=\,\, {\vec f}\times
\mbox{{\mathversion{bold}${\ell}$}}
\,\,+\,\, {\bf p}\times({\bf r}\times
{\vec f})\,.
\eeq
As we have seen above, the LRL always points towards the 
symmetry axis of the orbit. In the Kepler problem, 
for elliptical case, this symmetry axis is the major semi-axis 
of the orbit. Therefore, we can calculate the velocity of
precession of the orbit by simply computing the velocity 
of precession of the LRL vector. In the present work 
we are concerned with the quantum correction to the 
Newton gravitational potential. Such correction is 
expected to be a very small quantity, hence it is completely 
fair to employ a perturbative approach. With this consideration 
in mind, we compute the time average of the velocity of 
precession of the LRL vector for one period of the 
unperturbed orbit. One can recall that both 
$\mbox{{\mathversion{bold}${\ell}$}}$ and ${\bf A}$ 
are constants of motion for the unperturbed orbit. Therefore, 
for the sake of our calculation we can simply use 
$\mbox{{\mathversion{bold}${\ell}$}}$ and ${\bf A}$, instead 
of $\langle\mbox{{\mathversion{bold}${\ell}$}}\rangle$ and 
$\langle{\bf A}\rangle$. Taking, then, the time average 
of the time derivative of the LRL vector over the period of 
the unperturbed orbit, we obtain 
\beq
\label{LRLDifZeroMedia}
\left\langle\frac{d{\vec A}}{dt}\right\rangle =
 \left\langle {\bf f} \times
\mbox{{\mathversion{bold}${\ell}$}}\right\rangle +
 \left\langle{\bf p}\times({\bf r}\times
{\bf f})\right\rangle
= \mbox{{\mathversion{bold}${\Omega}$}}\times{\vec A}\, ,
\eeq
where ${\bf A}$ is the LRL 
vector in the unperturbed case, and we denoted by 
$\mbox{{\mathversion{bold}${\Omega}$}}$ 
the time average value of the velocity of precession of the 
LRL vector. Let us note that, in the {\l.h.s.} of the 
Eq. (\ref{LRLDifZeroMedia}), there is a quantity which 
is of the first order in the perturbing force. That is 
why the sign of averaging can not be omitted here. 

In Eq. (\ref{LRLDifZeroMedia}) the time-averaging of a function 
$F$ means
\beq
\langle F \rangle \,=\,
 \frac{1}{\tau}\,\int_0^{\tau}
  F\big[ r(t),\varphi(t)\big]\, dt\,,
\eeq
where $\tau$ is the period of the non-perturbed motion (for the 
closed orbit case. However, since we are mainly interested in the
trajectory, $r(\varphi)$, 
it is convenient to perform a change of variables and trade
the time integration for the angular integration, namely,
\beq
\label{MediaAngular}
\langle F \rangle  &=&
  \frac{m}{\ell\tau} \int_0^{2\pi} r^2(\varphi)
 F\big[ r(\varphi),\varphi\big] \, d\varphi\, .
\eeq

We note that for a central perturbative force ${\vec f}$, 
the angular momentum {\mathversion{bold}${\ell}$} is constant 
and the second term in the {\it r.h.s.} of Eq. (\ref{LRLDifZero}) 
vanishes. Then one can write, in the Cartesian basis,
\beq
\label{MediaForcaCentral}
\left\langle \frac{d {\vec A}}{dt} \right\rangle \,=\,
  \left\langle {\vec f} \times
 \mbox{{\mathversion{bold}${\ell}$}}\right\rangle
 \,\,=\,\,
 \langle f(r)\cos\varphi \rangle\, \hat{{\bf x}}
 \times \mbox{{\mathversion{bold}${\ell}$}}
 \,\,+\,\, \langle f(r)
 \,\sin\varphi\rangle\, \hat{{\bf y}}
 \times \mbox{{\mathversion{bold}${\ell}$}}\,.
\eeq

Since $ f$ depends only on the distance $r$ and the 
non-perturbed orbit is symmetric with respect to the 
major semi-axis, the second term in the {\it r.h.s.} 
of (\ref{MediaForcaCentral}) vanishes and we obtain
\beq
\label{PrecessaoForcaCentral}
\left\langle\frac{d{\bf A}}{dt}\right\rangle
\,=\,
{\mbox{{\mathversion{bold}${\Omega}$}}} \times
{\vec A}\,,
\quad \mbox{where} \quad
{\mbox{{\mathversion{bold}${\Omega}$}}} =
- \dfrac{\langle f(r)\cos\varphi \rangle}{mk\varepsilon}
\,\mbox{{\mathversion{bold}${\ell}$}}\,,
\eeq
describing the average precession of the orbit.

\section{\large\bf Solar system tests for logarithmic term}

Let us now apply the method developed in the previous section
to the case where the disturbing force is due to quantum 
effects. As we have already seen in Sect. 2, the corresponding 
additional term for the gravitational potential is given by
\beq
u(r) \,=\, \dfrac{m c^2}{2} \dfrac{\delta G}{G_0} \, ,
\eeq
where, according to Eqs. (\ref{RG4}) and (\ref{muset}),
we meet logarithmic dependence
\beq
G(\Phi_{\textrm{Newt}}) = \dfrac{G_0}{1\,+\,\nu\alpha\,
\ln \left(\Phi_{\textrm{Newt}}/\Phi_0\right)^2} \,.
\label{Main}
\eeq
From the previous equation, the perturbing force acting on 
the particle is given by 
\beq
\label{PertubacaoQuanticaCentral}
{\bf f} \,=\, -\,\dfrac{mc^2}{2G_0}\,\,
\mbox{grad}\, G(\Phi_{\textrm{Newt}})
\, =\,-\,\dfrac{m c^2\,\nu\al\,\hat{{\bf r}}}
{r \left[1\,+\,\nu\al\,\ln (r_0^2 / r^2)\right]^2} \,.
\eeq
From Eqs. (\ref{PrecessaoForcaCentral}) and  
(\ref{PertubacaoQuanticaCentral}), we obtain the 
following average precession rate,
\beq
\mbox{{\mathversion{bold}${\Omega}$}}
&=&
\frac{\nu\alpha c^2\,\mbox{{\mathversion{bold}${\ell}$}}}{k\vp}\,
\left\langle
\dfrac{\cos\varphi}{r}
\left[ 1+\nu\alpha\log\left(\dfrac{r_0^2}{r^2}\right)\right]^{-2}
\right\rangle
\cr\cr\cr &=&
\frac{\nu\al\, mc^2 }{\tau k\vp} \,
\, \frac{\mbox{{\mathversion{bold}${\ell}$}}}{\ell}
\,\int_0^{2\pi}
 \frac{r(\varphi) \cos\varphi}{\left[1
+ 2 \nu\alpha\,\ln (r_0/r) \right]^2}\, d\varphi\,.
\eeq

One can remember that already at the typical galaxy scale
$\nu\alpha \propto 10^{-7}$ and that the expected bound
for the Solar system should be essentially smaller. Therefore
it is justified to use an approximation  $\nu\alpha \ll 1$
and keep only first order terms in this parameter.
Then, after some simple calculations we arrive at the
following expression for the absolute value of the precession
velocity:
\beq
\Om 
&\simeq&
\frac{\nu\al \,m c^2}{\tau k\varepsilon} \,
\int_0^{2\pi} r(\varphi)\cos\varphi
\left[1-2\nu\alpha\ln\left(\dfrac{r_0^2}{r^2}\right)\right]\,
d\varphi
\label{formula}
\cr\cr
&\simeq&
\dfrac{\nu\al\,m c^2} {\tau k\varepsilon} \,
\int_0^{2\pi}
r(\varphi)\cos\varphi\, d\ph
\cr\cr
&=&
\dfrac{\nu\al\,m c^2 a(1-\vp^2)} {\tau k\vp}\,
\int_0^{2\pi} \dfrac{\cos\varphi}
{1+\varepsilon \cos \varphi}\, d\ph\,.
\eeq
The integral in the above expression can be easily solved
and as a result we find
\beq
{\Omega} \,=\,
\dfrac{2 \pi a c^2\,\,\al\nu
\,[\sqrt{1-\varepsilon^2}-(1-\varepsilon^2)]}
{G_0M \tau \varepsilon^2}\,.
\eeq
One can use Eq. (\ref {formula} )
for deriving an upper bound for the parameter $\al\nu$ in
the Solar system. For the case of the precession of Mercury, 
the uncertainty in the measurement of the velocity of 
precession is $0.45^{\prime\prime}$ per century.

After some simple calculations we arrive at the upper bound,
\beq
\nu \alpha < 10^{-17}\,,
\label{result}
\eeq
where we have used the following values:
\beq
G &=& 6.67 \cdot 10^{-11} \, Nm^2kg^{-2}
\,,\qquad
M = 1.98 \cdot 10^{30}\,kg
\,,\qquad
c = 3\cdot 10^{8} \,ms^{-2}\,,
\nonumber
\\
a &=& 6.97 \cdot 10^{10}m, \qquad 
\tau = 0.241\,\, \mbox{years}
\qquad \mbox{and} \qquad
\vp = 0.2056\,.
\label{numbers}
\eeq
Here $\tau,\,a,\,\vp$ are the period of rotation of Mercury,
the major semi-axis and eccentricity of its orbit. 

\section{\large\bf Conclusions.}

We have considered the running of the Newton constant $G$ in
the framework of a general Renormalization Group approach. The
covariance and dimensional arguments lead to a unique possible
form of such running, which can take place in all loop orders.
The beta-function for $G$ has one arbitrary parameter $\nu$
which depends on the details of the given theory. Vanishing
$\nu$ means there is no running at all, this means there are
no relevant quantum corrections into the low-energy sector
(Hilbert-Einstein) of the gravitational action.

Assuming that  $\nu$ is non-zero,
one can try to derive upper bound for $\nu$ from different
gravitational observations. In the recent paper \cite{RotCurve}
it was shown that the identification of the renormalization
group scale $\mu \sim (\Phi_{\rm Newt})^\al$ provides an
excellent fit for the rotation curves of the galaxies, with
the product $\al\nu$ being about $10^{-7}$ and moreover
steadily growing with the increase of the mass of the galaxy
under consideration. No Dark Matter is requested for
this fit. In the present work we  have derived an upper
bound for $\al\nu$ in the Solar system using the method 
based on the dynamics of the LRL vector. The maximal 
possible value we have obtained for $\al\nu$ is $10^{-17}$, 
which implies qualitatively the same form of a running that
was predicted in \cite{Hrvoje} on theoretical basis. 

Indeed, both values $\al\nu \propto 10^{-7}$ and  $10^{-17}$ 
have to be seen as maximal ones, representing upper bounds. 
In case of rotation curves one can admit certain amount of
a Dark Matter (with $\Om_{DM}^0$ smaller than usual), which
should be helpful to explain other observations (LSS, CMB, BAO
etc). Let us note that the first paper exploring the 
possibility of an alternative concordance model with taking 
into account quantum corrections is under preparation \cite{AlFa}.
In the case of a quasi-Newtonian potential in the Solar
system we can also see the value $10^{-17}$ as an upper bound,
such that the real value of  $\al\nu$ can be much smaller
than that. However, it is definitely remarkable that the
two different observations produced results which are
consistent with each other and also with the theoretical
prediction of \cite{Hrvoje}.

\section*{\large\bf Acknowledgments.}
Authors wish to thank Prof. Clifford Will for reading the 
preliminary version of the manuscript and making a useful 
remark on the presentation.  
C.F. and W.J.M. Kort-Kamp are grateful to CNPq, 
S.M. is grateful to CAPES and I.Sh. is grateful 
to CNPq, FAPEMIG and ICTP for partial financial support.


\end{document}